\documentstyle[12pt]{article}
\begin{document}

\def\ds{\displaystyle}
\def\beq{\begin{equation}}
\def\eeq{\end{equation}}
\def\bea{\begin{eqnarray}}
\def\eea{\end{eqnarray}}
\def\beeq{\begin{eqnarray}}
\def\eeeq{\end{eqnarray}}
\def\ve{\vert}
\def\vel{\left|}
\def\ver{\right|}
\def\nnb{\nonumber}
\def\ga{\left(}
\def\dr{\right)}
\def\aga{\left\{}
\def\adr{\right\}}
\def\lla{\left<}
\def\rra{\right>}
\def\rar{\rightarrow}
\def\nnb{\nonumber}
\def\la{\langle}
\def\ra{\rangle}
\def\ba{\begin{array}}
\def\ea{\end{array}}
\def\tr{\mbox{Tr}}
\def\ssp{{\Sigma^{*+}}}
\def\sso{{\Sigma^{*0}}}
\def\ssm{{\Sigma^{*-}}}
\def\xis0{{\Xi^{*0}}}
\def\xism{{\Xi^{*-}}}
\def\qs{\la \bar s s \ra}
\def\qu{\la \bar u u \ra}
\def\qd{\la \bar d d \ra}
\def\qq{\la \bar q q \ra}
\def\gGgG{\la g^2 G^2 \ra}
\def\q{\gamma_5 \not\!q}
\def\x{\gamma_5 \not\!x}
\def\g5{\gamma_5}
\def\sb{S_Q^{cf}}
\def\sd{S_d^{be}}
\def\su{S_u^{ad}}
\def\ss{S_s^{??}}
\def\sbp{{S}_Q^{'cf}}
\def\sdp{{S}_d^{'be}}
\def\sup{{S}_u^{'ad}}
\def\ssp{{S}_s^{'??}}
\def\sig{\sigma_{\mu \nu} \gamma_5 p^\mu q^\nu}
\def\fo{f_0(\frac{s_0}{M^2})}
\def\ffi{f_1(\frac{s_0}{M^2})}
\def\fii{f_2(\frac{s_0}{M^2})}
\def\O{{\cal O}}
\def\sl{{\Sigma^0 \Lambda}}
\def\es{\!\!\! &=& \!\!\!}
\def\ar{&+& \!\!\!}
\def\ek{&-& \!\!\!}
\def\cp{&\times& \!\!\!}
\def\se{\!\!\! &\simeq& \!\!\!}
\def\kpm{&\pm& \!\!\!}
\def\kmp{&\mp& \!\!\!}
\def\arr{\!\!\!\!&+&\!\!\!}
\def\eqv{&\equiv& \!\!\!}

\def\simlt{\stackrel{<}{{}_\sim}}
\def\simgt{\stackrel{>}{{}_\sim}}


\title{
         {\Large
                 {\bf
Magnetic moment of the $\rho$ meson in QCD light cone sum rules
                 }
         }
      }

\author{\vspace{1cm}\\
{\small T. M. Aliev \thanks
{e-mail: taliev@metu.edu.tr}\,\,,
\.{I} Kan{\i}k \thanks
{e-mail: e114288@metu.edu.tr}\,\,,
M. Savc{\i} \thanks
{e-mail: savci@metu.edu.tr}} \\
{\small Physics Department, Middle East Technical University, 
06531 Ankara, Turkey} }

\date{}

\begin{titlepage}
\maketitle
\thispagestyle{empty}

\begin{abstract}
   
The magnetic moment $\mu$ of the $\rho$ meson is studied in QCD light cone 
sum rules, and it is found that $\mu = (2.3 \pm 0.5)$. A comparison of 
our result on the magnetic moment of the $\rho$ meson with the predictions 
of the other approaches, is presented.

\end{abstract}

~~~PACS numbers: 11.50.Hx, 13.40.Em
\end{titlepage}

\section{Introduction}

QCD sum rules, which are based on the first principles of QCD \cite{R5501},
is a powerful tool in investigation of the hadron physics. In this method,
physically measurable quantities of hadrons are connected with QCD 
parameters, where hadrons are represented by their interpolating
quark current taken at large virtualities, and following that,  correlator 
of these quark currents is introduced. The main idea of the method
is to calculate the correlator with the help of operator product
expansion (OPE) in the framework of QCD (accounting for both perturbative
and nonperturbative contributions) and then connect them with the
phenomenological part. Physical quantities of interest are determined by
matching these two representations of the correlator. 

QCD sum rule method is successfully applied to many problems of hadron
physics (about the method see, for example, review papers
\cite{R5502}--\cite{R5505} and references therein).

One of the important static characteristic of hadrons is their magnetic 
moment. Magnetic moments of nucleons are calculated in the framework of 
the QCD sum rule method in \cite{R5506,R5507}, using the external field 
technique, and using the same approach magnetic moment of the $\rho$ 
meson is calculated in \cite{R5508}. 

Furthermore, it should be mentioned here that, in \cite{R5509} form 
factors of the $\rho$ meson are calculated at intermediate momentum transfer by 
using the three--point QCD sum rules method, and then extrapolating
these form factors to $Q^2=0$ (this point lies outside the
applicability region of the method).

In this work, we present an independent calculation of the magnetic moment
of the $\rho$ meson in the framework of an alternative approach to the
traditional QCD sum rules, i.e., QCD light cone sum rules method (QLCSR).

Few words about this method are in order. The QLCSR method is based on
Operator Product Expansion (OPE) near light cone, which is an expansion
over the twist of the operators rather than dimensions as in the traditional
QCD sum rules. The nonperturbative dynamics encoded in the light cone wave
functions, determines the matrix elements of the nonlocal operators between
the vacuum and the hadronic states (more about this
method and its applications can be found in \cite{R5505,R5510})

The QLCSR is successfully applied to a variety of problems in hadron
physics. For example, magnetic moments of the octet and decuplet baryons are
calculated in \cite{R5511} and \cite{R5512}, respectively, and magnetic
moment of the nucleon is first obtained in QLCSR in \cite{R5513}.

The paper is organized as follows. In section 2, QLCSR for the $\rho$ meson
magnetic moment is obtained. In section 3, our numerical results and a
comparison with the results of the other approaches is presented.

\section{QLCSR for the $\rho$ meson magnetic moment}

In this section we calculate the $\rho$ meson magnetic moment in QLCSR. We
consider the following correlator of two vector currents in the external
electromagnetic field

\bea
\label{e5501}
\Pi_{\mu\nu} (p,q) = i \int d^4x e^{ipx} \lla 0 \vel \mbox{\rm T} \{
j_\nu(x) j_\mu^\dagger (0) \} \ver 0 \rra_\gamma ~,
\eea
where the subscript $\gamma$ denotes the external electromagnetic field, 
$j_\nu (x) = \bar{u} \gamma_\nu d(x)$ is the vector current with the 
$\rho$ meson quantum number.

Firstly, let us calculate the phenomenological part of the correlator.
By inserting a complete set of states between the currents in
Eq. (\ref{e5501}) with quantum numbers of the $\rho$ meson, we obtain the
following representation of the correlator
\bea
\label{e5502}
\Pi_{\mu\nu} = \frac{\la 0 \ve j_\nu \ve \rho(p) \ra
                     \la \rho(p) \ve \rho(p^\prime) \ra_\gamma
                     \la \rho(p^\prime) \ve j_\mu^\dagger\ve 0 \ra}
                    { ( p^2-m_\rho^2 ) ( p^{\prime 2} - m_\rho^2 )}
                + \cdots ~,
\eea 
where $p^\prime =p+q$, $q$ is the photon momentum and $\cdots$ describe
higher states and continuum contributions. The matrix element
$\la 0 \ve j_\nu \ve \rho(p) \ra$ is determined as
\bea
\label{e5503}
\la 0 \ve j_\nu \ve \rho(p) \ra = \frac{m_\rho^2}{g_\rho} 
\varepsilon_\nu (p)~.
\eea

Assuming parity and time--reversal invariance, the electromagnetic vertex of
the $\rho$ meson can be written in terms of three form factors \cite{R5514}
\bea
\label{e5504}
\la \rho(p,\varepsilon^r) \ve \rho(\rho^\prime,\varepsilon^{r^\prime})
\ra_\gamma \es - \varepsilon^\rho (\varepsilon^r)^\alpha
(\varepsilon^{r^\prime})^\beta \Big\{ G_1(Q^2) g_{\alpha\beta}
(p+p^\prime)_\rho + G_2(Q^2) (q_\alpha g_{\rho\beta} - q_\beta
g_{\rho\alpha}) \nnb \\
\ek \frac{1}{2 m_\rho^2} G_3(Q^2) q_\alpha q_\beta (p+p^\prime)_\rho
\Big\}~,
\eea
where $\varepsilon_\rho$ is the photon and $(\varepsilon^r)^\alpha$,
$(\varepsilon^{r^\prime})^\beta$ are the $\rho$ meson vector polarizations.
The Lorentz invariant form factors $G_i(Q^2)$ are related to the charge,
magnetic and quadropole form factors through the relations
\bea
\label{e5505}
F_C \es G_1 +\frac{2}{3} \eta F_{\cal D}~, \nnb \\
F_M \es G_2~, \nnb \\
F_{\cal D} \es G_1 - G_2 + (1+\eta) G_3~,
\eea
where $\eta = Q^2/4 m_\rho^2$ is a kinematical factor.
At zero momentum transfer, these form factors are proportional to the usual
static quantities of charge, magnetic moment $\mu$ and quadropole moment
${\cal D}$:
\bea
\label{e5506}
e F_C(0) \es e~, \nnb \\
e F_M(0) \es 2 m_\rho \mu~, \nnb \\
e F_{\cal D}(0) \es m_\rho^2 {\cal D}~.
\eea

Using Eqs. (\ref{e5502})--(\ref{e5504}) and performing summation over
polarizations of the $\rho$ meson, for the phenomenological part 
of the correlator we get
\bea
\label{e5507}
\Pi_{\mu\nu} \es \frac{m_\rho^4}{g_\rho^2} \varepsilon^\rho
\frac{1}{(m_\rho^2-p^2) [m_\rho^2-(p+q)^2]} \nnb \\
\cp \Bigg\{ G_1(Q^2) (p+p^\prime)_\rho \Bigg[g_{\mu\nu} -
\frac{p_\mu p_\nu}{m_\rho^2} - \frac{p_\mu^\prime p_\nu^\prime}{m_\rho^2}
+\frac{ p_\mu^\prime p_\nu}{2 m_\rho^4} (Q^2+2 m_\rho^2) \Bigg] +
G_2(q^2) \Bigg[q_\mu g_{\nu\rho} \nnb \\
\ek  q_\nu g_{\mu\rho}
- \frac{ p_\nu}{m_\rho^2} \ga q_\mu p_\rho - \frac{1}{2} Q^2 g_{\mu\rho} \dr
+\frac{ p_\mu^\prime}{m_\rho^2} \ga q_\nu p_\rho^\prime + \frac{1}{2} Q^2
g_{\nu\rho} \dr -
 \frac{ p_\mu^\prime p_\nu p_\rho }{m_\rho^4}
Q^2\bigg] \nnb \\
\ek \frac{1}{2 m_\rho^2} G_3(Q^2) (p+p^\prime)_\rho \Bigg[q_\mu q_\nu -
\frac{p_\nu q_\mu}{m_\rho^2} \frac{1}{2} Q^2 +
\frac{p_\mu^\prime q_\nu}{m_\rho^2} \frac{1}{2} Q^2 -
\frac{p_\nu p_\mu^\prime}{m_\rho^4} \frac{1}{4}(Q^2)^2 \Bigg] \Bigg\}~,
\eea
where $Q^2=-q^2$. Throughout our analysis, only the values of the form factors at
$Q^2=0$ are needed. Additionally, using $p^\prime=p+q$ and $q\varepsilon=0$,
Eq. (\ref{e5507}) can be simplified and final answer for the
phenomenological part can be written as
\bea
\label{e5508}
\Pi_{\mu\nu} \es \frac{m_\rho^4}{g_\rho^2} \frac{\varepsilon^\rho}
{[m_\rho^2-(p+q)^2]} \Bigg\{2 p_\rho F_C(0) \Bigg[g_{\mu\nu} -
\frac{p_\mu p_\nu}{m_\rho^2}- \frac{p_\mu q_\nu}{m_\rho^2} \Bigg] \nnb \\
\ar F_M(0) \Bigg[q_\mu g_{\nu\rho}- q_\nu g_{\mu\rho} +
\frac{1}{m_\rho^2} p_\rho (p_\mu q_\nu- p_\nu q_\mu) \Bigg]
- \Big(F_C(0)+F_{\cal D}(0)\Big) \frac{p_\rho}{m_\rho^2} q_\nu q_\mu \Bigg\}~.
\eea
In order to extract out the magnetic moment of the $\rho$ meson from Eq.
(\ref{e5508}), we will chose the structure
$(p\varepsilon)(p_\mu q_\nu-p_\nu q_\mu)$. Hence, the phenomenological part of
the correlator for the above--mentioned structure can be written as
\bea
\label{e5509}
\Pi = \frac{m_\rho^2}{g_\rho^2} \frac{1}{(m_\rho^2-p^2) [m_\rho^2-(p+q)^2]}
\, \mu~,
\eea
where $\mu$ is the $\rho$ meson magnetic moment in units of $e/2m_\rho$. 

Our next task is calculation of the correlator in Eq. (\ref{e5501})
from the QCD side. The correlator receives perturbative and nonperturbative
contributions. The perturbative part corresponds to on--shell 
photon emission from virtual quarks and it is described by the triangle 
diagram (see Fig. (1)). 
In order to calculate the nonperturbative contributions (see Fig. (2)), 
we need the
matrix elements of the nonlocal operators between the vacuum and the photon
states, i.e., $\la \gamma(q) \ve \bar{q}(x) \Gamma_i (0) \ve 0\ra$, where
$\Gamma$ is an arbitrary Dirac matrix. In our calculations we take into
account twist--2, 3 and 4 photon wave functions (more about the photon wave
functions, see \cite{R5515}). In what follows we present definitions whose 
wave functions give contribution only to the structure $(p\varepsilon)
(p_\mu q_\nu - p_\nu q_\mu)$.
\bea
\label{e5510}
\lla \gamma (q) \ve \bar{q} (x) \gamma_\mu q(0) \ve 0 \rra \es
e e_q f_{3\gamma} \ga \varepsilon_\mu - q_\mu \frac{\varepsilon x}{qx} \dr
\int_0^1 du e^{iuqx} \psi^{(v)} (u) ~, \\ \nnb \\
\label{e5511}
\lla \gamma (q) \ve \bar{q} (x) \gamma_\mu \gamma_5 q(0) \ve 0 \rra \es
- \frac{1}{4} e e_q f_{3\gamma} \epsilon_{\mu\alpha\beta\rho}
\varepsilon^\alpha q^\beta x^\rho \int_0^1 du e^{iuqx} \psi^{(a)} (u)~, \\ \nnb \\
\label{e5512}
\lla \gamma (q) \ve \bar{q} (x) \sigma_{\mu\nu} q(0) \ve 0 \rra \es
-i e e_q \la \bar{q}q \ra (\varepsilon_\mu q_\nu - \varepsilon_\nu q_\mu)
\int_0^1 du e^{iuqx} \Bigg\{\chi \phi_\gamma (u) + \frac{x^2}{16} A (u)
\Bigg\} \nnb \\
- i e e_q \la \bar{q}q \ra \Bigg[ x_\nu \Bigg( \varepsilon_\mu \!\!\!\!\ek q_\mu 
\frac{\varepsilon x}{qx} \Bigg) -
x_\mu \ga \varepsilon_\nu - q_\nu \frac{\varepsilon x}{qx} \dr \Bigg]
\int_0^1 du e^{iuqx} h_\gamma (u)~,
\eea
where $\phi_\gamma (u)$ is twist--2, $\psi^{(v)}(u)$ and $\psi^{(a)}(u)$ are
twist--3, $A(u)$ and $h_\gamma (u)$ are twist--4 photon wave functions,
respectively, and $\chi$ is the magnetic susceptibility. It should be noted
here that, there are several other functions $T_i(\alpha_i)$ and 
$\widetilde{S}(\alpha_i)$ (for their definitions, see \cite{R5515}) that also 
give contribution to the above--mentioned structure. But their contributions 
are proportional to the quark mass (in our case $u$ and $d$ quark masses) 
and therefore irrelevant in the massless quark case.

After some effort, we get the following expression for the correlator from
QCD side in the $x$--representation 
\bea
\label{e5513}
\Pi_{\mu\nu} \es e \int_0^1 du \int dx e^{i(p+uq)x} \varepsilon x (x_\mu
q_\nu - x_\nu q_\mu ) \nnb \\
\cp \Bigg\{ (e_d-e_u) \Bigg[ \frac{3}{4 \pi^4 x^6} - f_{3\gamma}
\frac{\psi^{(a)}(u)}{8 \pi^2 x^4} + \frac{i}{2} f_{3\gamma}
\frac{\psi^{(v)}(u)}{\pi^2 (qx) x^4} -
\frac{m_0^2}{384 (qx)} \la \bar{q}q \ra^2 h_\gamma (u) \Bigg] \nnb \\
\ek \frac{(e_d+e_u)}{8 (qx)} h_\gamma (u) \Bigg\}~.
\eea 
  
Using Eq. (\ref{e5513}) and after performing Fourier transformation, the
result for the structure $(p \varepsilon)(q_\nu p_\mu - q_\mu q_\nu)$ can
be obtained. The sum rules for the $\rho$ meson can be obtained after applying
double Borel transformation on the variables $p^2$ and $(p+q)^2$, which
suppresses the continuum and higher states contributions (about this
procedure, see \cite{R5511,R5512,R5516,R5517}, and references therein) and
then matching both representations of the correlators.

Finally, for the above--mentioned structure we get the following sum rule for 
the $\rho$ meson magnetic moment
\bea
\label{e5514}
\mu = \frac{g_\rho^2}{m_\rho^2} e^{m_\rho^2/M^2} (e_u-e_d) 
\Bigg\{ \frac{3}{8 \pi^2} M^2 f_0(s_0/M^2) + \frac{f_{3\gamma}}{2}
\psi^{(a)}(u_0) - 2 f_{3\gamma}\Psi^{(v)}(u_0)\Bigg\}~,
\eea
where 
\bea
\Psi^{(v)} (u) = \int_0^u \psi^{(v)} (v) dv~, \nnb
\eea
and, the function
\bea
f_0(s_0/M^2) = 1-e^{-s_0/M^2}~,\nnb
\eea
is used to subtract continuum contributions, and naturally, the Borel parameters 
$M_1^2$ and $M_2^2$ are set to be equal to each other, i.e., $M_1^2=M_2^2\equiv 
2 M^2$ since we are dealing with just a single meson, and hence
\bea
u_0 = \frac{M_1^2}{M_1^2+M_2^2} = \frac{1}{2}~.\nnb
\eea
Note that, the last two terms in Eq. (\ref{e5513}) disappear after double
Borel transformation is performed.

The main reason why we choose the structure $(p \varepsilon)(q_\nu p_\mu -
q_\mu q_\nu)$ is that, the term proportional to the magnetic susceptibility
$\chi$ does not give any contribution, and hence the main uncertainty coming from
the definition of $\chi$ is absent in the sum rule.

\section{Numerical analysis}

In this section we present our numerical analysis on the $\rho$ meson
magnetic moment. It follows from Eq. (\ref{e5514}) that, in order to perform
further numerical analysis one needs to know the photon wave functions
$\psi^{(a)}(u)$ and $\psi^{(v)}(u)$. The explicit expressions of the
functions are
\bea
\psi^{(v)} (u) \es 10 u (1-3 u+2 u^2) - \frac{15}{8} u (w_\gamma^A -
3 w_\gamma^V) (1 - 10 u + 30 u^2 - 35 u^3 + 14 u^4) ~, \nnb \\
\psi^{(a)} (u) \es \frac{5}{2} \Bigg[ 1 +\frac{9}{16} w_\gamma^V -
\frac{3}{24} w_\gamma^A \Bigg] [1-(2u-1)^2] [5 (2u-1)^2-1]~.\nnb
\eea

The values of the input parameters $w_\gamma^V$, $w_\gamma^A$ and
$f_{3\gamma}$ are given in \cite{R5515} to have the values: $w_\gamma^V=(3.8 \pm
1.8)$, $w_\gamma^A=-(2.1\pm 1.0)$ and $f_{3\gamma}=-(3.9
\pm 2.0)\times10^{-3}~GeV^{-2}$. The remaining input parameters are $m_\rho=
0.77~GeV$ and $g_\rho^2/4 \pi  = 1.27$.

In Fig. (3) we present the dependence of the magnetic moment on $M^2$ at
three different values of the continuum threshold: $s_0=1.5~GeV^2$,
$s_0=1.8~GeV^2$ and $s_0=2.0~GeV^2$. Note that, $M^2$ in the sum rule is an
auxiliary parameter and the physical quantities are expected to be
independent of it. Therefore, one must look for a region of $M^2$ for which the
magnetic moment $\mu$ be practically independent of it. The lower limit of
$M^2$ is determined by the requirement that terms $\sim M^{-2n}~(n>1)$
remain subdominant. In other words, large power corrections must be absent
in the sum rule. The upper bound of $M^2$ is determined by demanding that
the contributions of the higher resonances and continuum are less than, for
example, $30\%$ of the total result. Our numerical calculation shows that
these requirements are satisfied in the region $1.0~GeV^2 \le M^2 \le
1.4~GeV^2$ and magnetic moment in this region is practically independent of
$M^2$. We also see from this figure that as $s_0$ varies from
$s_0=1.5~GeV^2$ to $s_0=2.0~GeV^2$, the magnetic moment of the $\rho$ meson 
changes by an amount of approximately $10\%$. Therefore we can conclude
that the result seems to be almost insensitive to the change in $s_0$ and
$M^2$ in the above--mentioned region. The final result for the magnetic
moment of the $\rho$ meson turns out to be 
\bea
\mu = 2.3 \pm 0.5~,\nnb
\eea
in units of $(e/2m_\rho)$, where the error can be attributed to the
variations in $s_0$, $M^2$ and uncertainties in the values of $f_{3\gamma}$,
$w_\gamma^V$ and $w_\gamma^A$.

At the end, we would like to present a comparison of our result on the $\rho$
meson magnetic moment, with the ones existing in literature. In the
Dyson--Schwinger based models, the $\rho$ meson magnetic moment is estimated
to have the value $\mu = 2.69$ \cite{R5518},  $2.5 \le \mu \le 3$
\cite{R5519}
in units of $e/2 m_\rho$. Covariant and Light front approaches with 
constituent quark model, both, predict $\mu = 2.23 \pm 0.13$ \cite{R5520}
and in light front formalism it is estimated to be $\mu=1.83$ \cite{R5521}.
The magnetic moment of $\rho$ meson was calculated long time ago in
\cite{R5522}, by
considering the low energy limit of the radiative amplitudes in conjunction
with the amplitude calculated by the hard--pion technique and found that
\bea
\frac{16\pi^2 \alpha^2 g_\rho^2}{m_\rho^2 \int ds \sigma_{e^+ e^- \rar n}}
<\mu_\rho < 2~. \nnb
\eea
The $\rho$ meson magnetic moment was also calculated in lattice theory which
predicted $\mu_\rho=2.25(34)$ \cite{R5523}.  
As has already been noted, the magnetic moment of the $\rho$ meson in the
framework of the traditional QCD sum rule in the presence of external field,
is calculated in \cite{R5508} and it is obtained that
$\mu = 1.5 \pm 0.3$. Our result is closer to the predictions of the works
\cite{R5520} and \cite{R5523}.   
  
Finally, we would like to discuss briefly the question how to measure the
magnetic moment of $\rho$ meson in experiments. At present, even upper bound
for the magnetic and quadropole moments of $\rho$ meson are absent. The very
short lifetime does not allow the use of vector--meson--electron scattering or
spin procession technique \cite{R5524} to measure the above--mentioned
quantities.

An alternative method for determination of the multipole moments of particles
is based on soft photon emission off the hadrons was proposed in
\cite{R5525}, since the photon carries information on higher multipoles of
the emitting particles. The main idea of this work is that the amplitude for
radiative process can be expressed as a power expansion in the photon energy
$w$ as follows
\bea
M = \frac{A}{w} + B w^0 + C w + \cdots \nnb
\eea
The electric charge contribute to the amplitude at order $w^{-1}$ and the
contribution coming from magnetic moment is proportional to $w^0$.
Therefore, by measuring the cross section or decay width of the radiative
process and neglecting terms linear in $w$, one can determine the magnetic
moments of charged particles.  

In \cite{R5525} and \cite{R5526}, the possibility of measuring the magnetic 
moment of the charged $\rho$ meson in radiative production and decays of 
such mesons are mentioned and it is claimed that, combined angular and energy
distributions of radiated photons is an efficient tool in measuring the
magnetic moment of the charged $\rho$ meson.

\section*{Acknowledgments}
We sincerely thank A. \"{O}zpineci for the useful help he provided,
and for his valuable comments and suggestions.

\newpage

\section*{Figure captions}
{\bf Fig. (1)} Diagrams describing perturbative contribution to the
correlator in Eq. (\ref{e5501}).\\ \\
{\bf Fig. (2)} Diagrams describing nonperturbative contribution to the
correlator in Eq. (\ref{e5501}). Here, Fig. (2a) corresponds to the leading
order contribution and Fig. (2b) corresponds to the gluon correction to the
correlator in Eq. (\ref{e5501}). In these figures, the wavy line represents
gluon, and solid lines represent quark fields, respectively.\\ \\   
{\bf Fig. (3)} The dependence of the magnetic moment of the $\rho$ meson on 
the Borel parameter $M^2$, at three different values of the continuum
threshold; $s_0=1.5~GeV^2$, $s_0=1.8~GeV^2$ and $s_0=2.0~GeV^2$.

\newpage

\begin{figure}
\vskip 1.5 cm
    \includegraphics{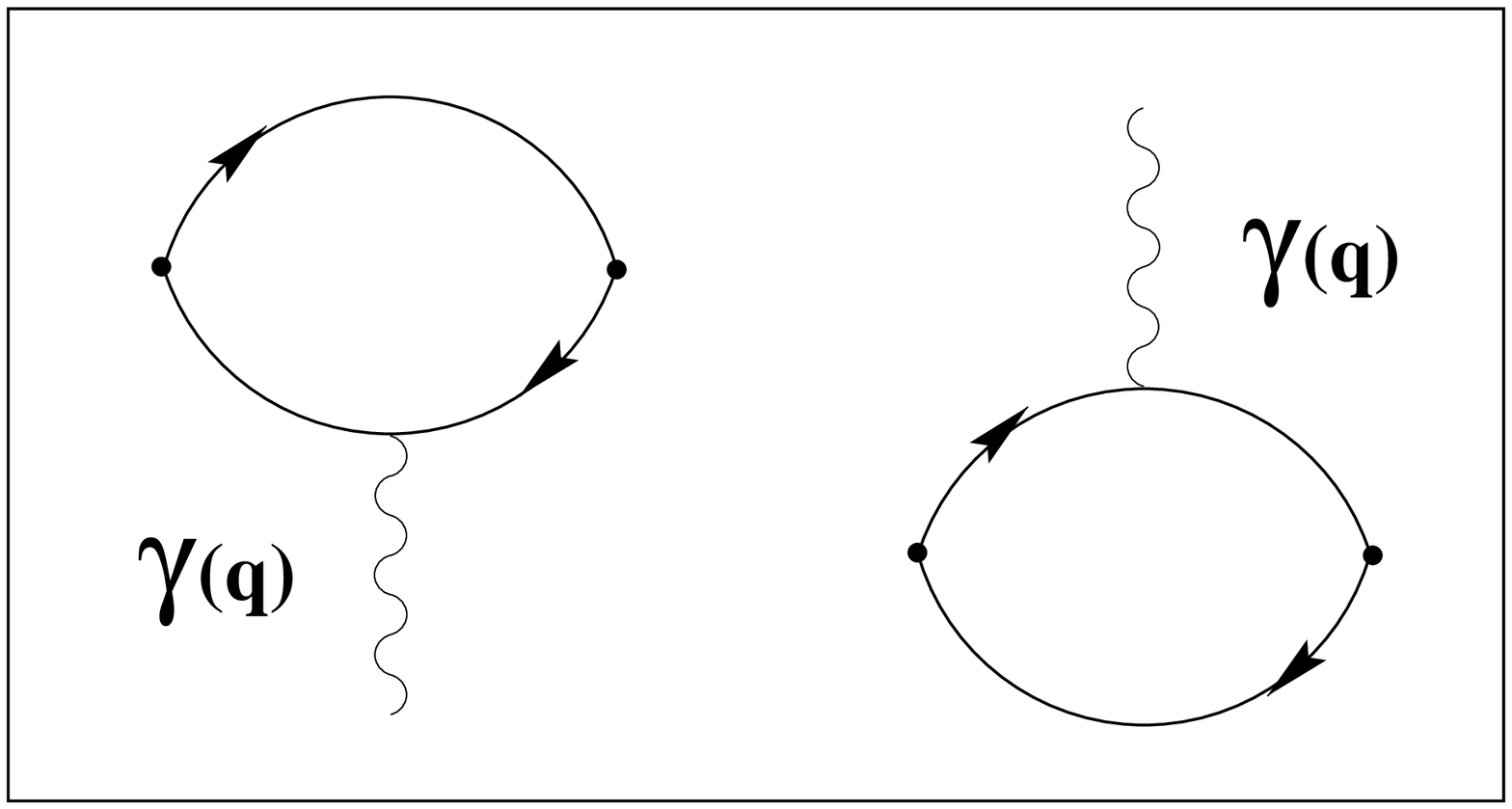}
\vskip 7.8cm
\caption{}
\end{figure}

\begin{figure}
\vskip 2.5 cm
    \includegraphics{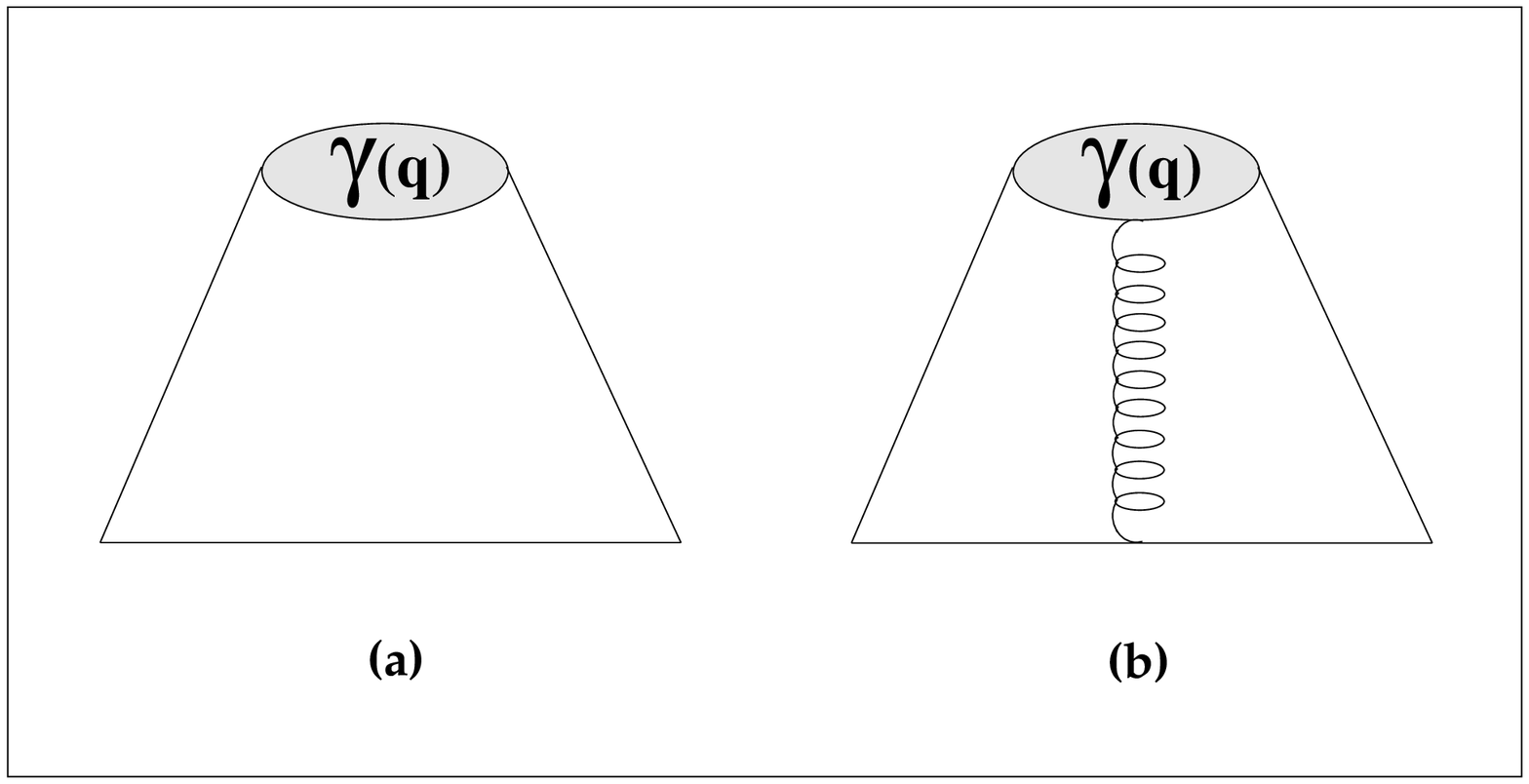}
\vskip 7.8 cm
\caption{}
\end{figure}

\begin{figure}
\vskip 1.5 cm
    \includegraphics{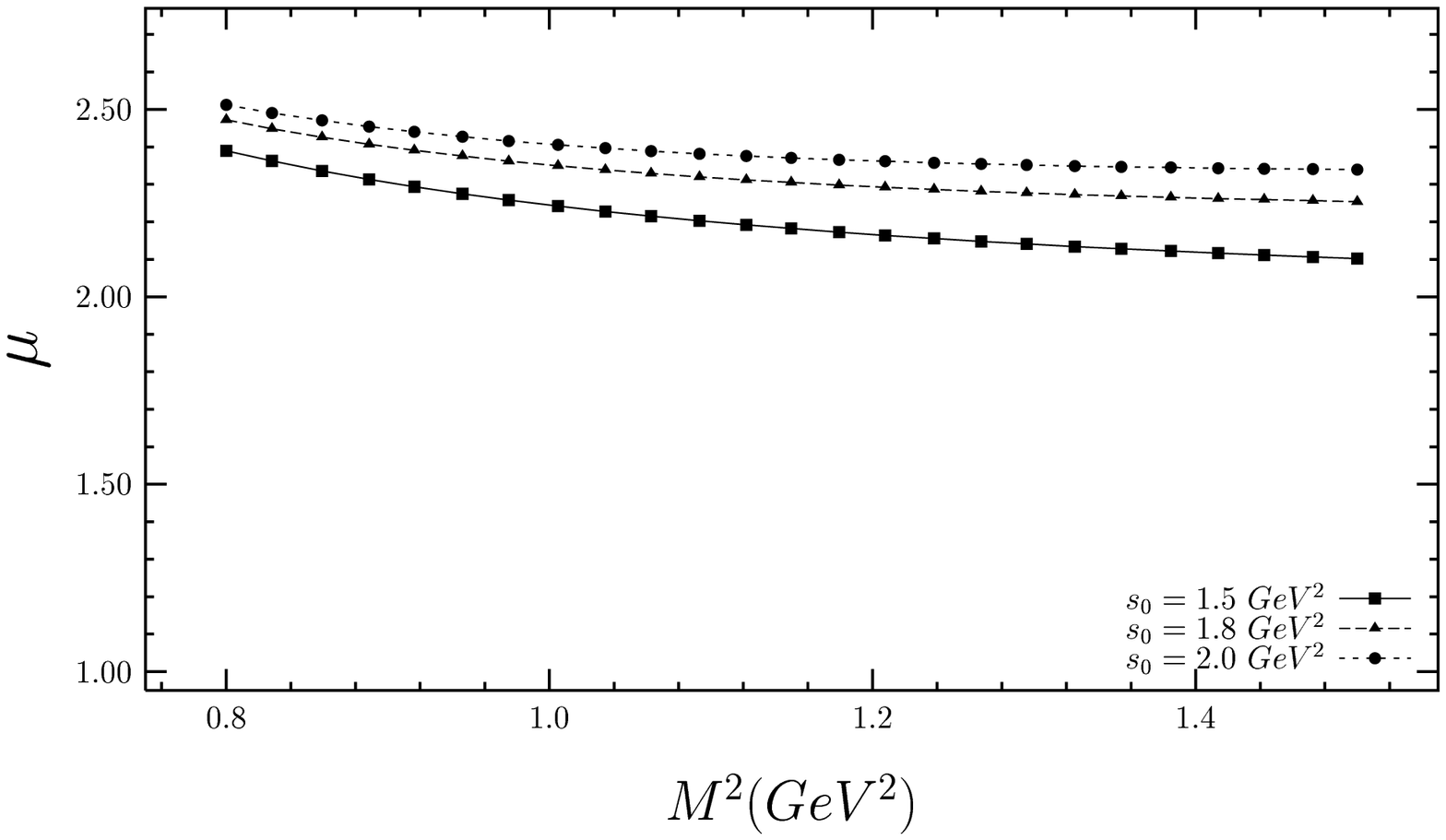}
\vskip 7.8cm
\caption{}
\end{figure}

\end{document}